\documentclass[10pt]{article}

\usepackage{cite} 

\usepackage{amsmath}
\usepackage{amssymb}
\usepackage{authblk}
\usepackage{array}
\usepackage{array} 
\usepackage{pifont}
\usepackage[usenames, dvipsnames]{color}
\usepackage{tabu}  
\usepackage{graphicx}
\usepackage{dcolumn} 
\usepackage{bm}      

\usepackage{soul} 

\usepackage{geometry}
\usepackage{multicol}
\usepackage{color}
\newcommand{\rv}[1]{{\color{Black}#1}}

\usepackage{graphicx}
\usepackage{tikz}
\usepackage{todonotes} 
\newcommand*\circled[1]{\tikz[baseline=(char.base)]{
            \node[shape=circle,draw,inner sep=0.5pt] (char) {#1};}}

\title{\bf \rv{A unified formal framework for developmental and evolutionary change in gene regulatory network models}}

\author[1,2]{Enrico Borriello\thanks{enrico.borriello@asu.edu}} 
\author[1,2,5,7]{Sara I. Walker\thanks{sara.i.walker@asu.edu}}
\author[1,3,4,6]{Manfred D. Laubichler\thanks{manfred.laubichler@asu.edu}}

\affil[1]{ASU-SFI Center for Biosocial Complex Systems, Arizona State University,Tempe, AZ, USA} 
\affil[2]{Beyond Center for Fundamental Concepts in Science, Arizona State University, Tempe AZ}
\affil[3]{Santa Fe Institute, Santa Fe, NM, USA}
\affil[4]{Marine Biological Laboratory, Woods Hole, MA, USA}
\affil[5]{School of Earth and Space Exploration, Arizona State University, Tempe AZ}
\affil[6]{School of Life Sciences, Arizona State University, Tempe AZ}
\affil[7]{Blue Marble Space Institute of Science}

\begin{document}

\maketitle


\begin{changemargin}{1cm}{1cm}  
\noindent {\bf Abstract:}
The two most fundamental processes describing change in biology, development and evolution, occur over drastically different timescales, difficult to reconcile within a unified framework. Development involves temporal sequences of cell states controlled by hierarchies of regulatory structures. It occurs over the lifetime of a single individual, and is associated to the gene expression level change of a given genotype. Evolution, by contrast entails genotypic change through the \rv{acquisition/loss of genes and changes in the network topology of interactions among genes. It} involves the emergence of new, environmentally selected phenotypes over the lifetimes of many individuals. Here we present a model of regulatory network evolution that accounts for both timescales. We extend the framework of \rv{Boolean} models of gene regulatory networks (GRN)-currently only applicable to describing development-to include evolutionary processes. As opposed to one-to-one maps to specific attractors, we identify the phenotypes of the cells as the relevant macrostates of the GRN. A phenotype may now correspond to multiple attractors, and its formal definition no longer requires a fixed size for the genotype. This opens the possibility for a quantitative study of the phenotypic change of a genotype, which is itself changing over evolutionary timescales.  We show how the realization of specific phenotypes can be controlled by gene duplication events \rv{(used here as an archetypal evolutionary event able to change the genotype)}, and how successive events of gene duplication lead to new regulatory structures via selection. \rv{At the same time, we show that our generalized framework does not inhibit network controllability and the possibility for network control theory to describe epigenetic signalling during development.}\\[2mm]
\end{changemargin}

\begin{multicols}{2}

\section{Introduction} 

Understanding the mechanisms underlying the emergence and persistence of new cell types is a central problem in the evolution and development of multicellular organisms.
Whereas all cell types can in principle access the same genetic information, in practice, regulation of gene expression restricts this such that only a subset of an organism's total genomic information content is accessible to a given cell type at a given time, permitting differentiation of many phenotypes from a single genotype \cite{Slack2013}. Regulation of gene expression therefore plays a dominant role in establishing cell types. From a formal point of view, the question of how new cell types emerge, therefore reduces to the problem of understanding how new regulatory structures evolve that can specify and control the expression of novel phenotypes.

The interplay among these regulatory genes, and their interaction with the other components of the cell governs the expression levels of both mRNA and proteins, where the set of interactions are described as a {\it gene regulatory network} (GRN). 
\rv{
If multicellular phenotypes are the product of developmental differentiation processes controlled by GRNs, as the raw material for evolutionary change, every phenotypic variant is expected to be the product of a corresponding change in those regulatory networks. 

In the case of evolution, continuous feedback in the form of selection or drift\cite{Lynch2007} leads to changes in the underlying network architecture controlling individual phenotypes, for instance by means of mutation, gene duplication or deletion, etc.
In the case of developmental differentiation, continued extracellular feedback over the regulatory network assures that the steady gene activation patterns (i.e. the ID of the cell) follow a specific developmental sequence. Such extracellular signalling is not required to change the underlying structure of the GRN for steady activation pattern to change.    
Nonetheless, this signalling induced change can have the same phenotypic effect as a structural change to the network (as during evolutionary change). 
In this manuscript, we examine the possibility of achieving a common description of both evolutionary change and developmental processes in GRN models.

When a GRN is described with the language of dynamical systems, extracellular feedback becomes formally equivalent to the notion of network controllability, and the interested reader is redirected to the excellent book of Iglesias and Ingalls for an extensive review of the subject \cite{Iglesias2010}.

Things change drastically when the evolutionary timescale is involved.} Any attempt at trying to predict the steady states of the regulatory process --i.e. the cell types-- in terms of a dynamical model (coupled ordinary differential equations, \rv{Boolean} network models, stochastic gene networks, to name a few common approaches) faces the difficulty of having to reconcile the fixed number of genes in these models, whose expression level is representative of a given cell type \cite{Stegle2015,Shapiro2013,Schwartzman2015}, with the possibility for the size of the genotype to change over evolutionary timescales.
Chromosome loss and gain, single gene and whole genome duplication, as well as horizontal gene transfer all alter the number of genes participating in the dynamics of a GRN. In doing so, these processes deprive the mapping of cell types to gene expression patterns of its original meaning. Therefore, even when successful in explaining developmental change, current GRN models must be redefined after each modification of the genotype for their use in evolutionary biology.


In this manuscript, we generalize the well established framework of \rv{Boolean}, dynamical models of GRNs as proposed by Kauffman \cite{Kauffman1969} \rv{--based on the hypothesis that cell types represent the attractor states of the GRN dynamics \cite{Delbruck1949,Jacob1961}--} to include features of evolutionary biology. 
In \rv{\rv{Boolean}} models, the attractors of the network dynamics encode different, stable cellular phenotypes, permitting a model for how multiple cell type identities can be encoded in the same regulatory structure.  
The novelty of our approach consists in relaxing the restrictive one-to-one mapping between network attractors and cell phenotypes by redefining phenotypes as {\it collections of gene expression patterns} with a given subset of genes sharing the same pattern (section 2). While the traditional definition, assuming a one-to-one map between phenotype and genotype, yields increasingly fine-tuned specifications for the phenotype for progressively larger genotypes, our novel definition \rv{identifies} the phenotypes with a {\it macrostate}, as opposed to individual (micro)states, of a dynamical system, and as such does not require fine-tuning.  

We will show that, under the relaxed assumption of identifying phenotypes as macrostates of the underlying \rv{\rv{Boolean}} GRN, a fixed genotypic size is not necessary for specifying or retaining phenotypes through evolutionary processes. We will exploit this possibility to study the emergence of new cell types, as well as the consolidation or loss of old types \rv{Section 2}, as a consequence of the changing size and topology of the GRN over evolutionary timescales, and of shifting environmental conditions. As such, our model also addresses an inconsistency arising from considering concepts belonging to different levels of description of a well conceived ontology of biological objects \cite{Laubichler2017} as being modeled as {\it same-level} processes: for example, gene expression levels and phenotypes, as they were interchangeable. Our approach will instead assume gene expression to be at a lower level of the ontology than phenotype, while phenotype and environment will belong to the same, higher ontological level.

In what follows, we focus on the case study of gene duplication, and use it as an example of genotype-changing evolutionary process.
Gene duplication has occurred in all three domains of life \cite{Kondrashov2002}, and is an ancient mechanism dating to before the last universal common ancestor of all life on Earth \cite{Conant2008}. \rv{Most genomic evolutionary processes include at least some gene duplications. It} is, by far, the dominant force in creating new genes, and at least 50\% of  genes in prokaryotes \cite{Brenner1995,Teichmann1998} and over 90\% of those in eukaryotes \cite{Gough2001} are the result of gene duplication. Nonetheless, with the exception of a few papers \cite{Wagner1994,Aldana2007,Crombach2008}, it has rarely been discussed as a mechanism for evolving new regulatory patterns for cell type identity.

\rv{Our preliminary study seems to suggest} an interplay between gene duplication and natural selection as the driving force responsible for the assemblage of genetic {\it modules}, or {\it core sets of regulatory genes}
\cite{Bonner1988,Wagner1996, Alberts1998,Hartwell1999,Pereira-Leal2006,Pereira-Leal2007,Achim2014}.
Evidence in favor of modularity in biology has steadily increased over the last three decades, and, as of today, modules have been both reconstituted {\it in vitro} 
\cite{Khalil2010}, 
and transplanted from one cell type to another \cite{Hsu1993}. 
It is now well accepted that biological functions are only rarely attributed to individual molecules - the role of hemoglobin in transporting oxygen along the bloodstream being among the best examples. Far more often, a biological function results instead from the interaction among many different proteins, like in the transduction process converting pheromone detection into the act of mating in yeast \cite{Stock1996,Herskowitz1995,Posas1998}. Network controllability is usually reinterpreted in GRN dynamical models as the mathematical  counterpart of extracellular signaling. Our macrostate interpretation of the phenotypes still allows us to adopt and assign biological meaning to network controllability techniques. For example, the control kernel of a GRN is defined in \cite{Kim2013} as the minimum number of genes/nodes whose expression it is necessary to regulate to steer the dynamics of the rest of the network toward a desired attractor (phenotype), {\it e.g.} an attractor associated with a functional cellular phenotype. Focusing on the developmental timescale we show it is still possible to easily rephrase, generalize, and adapt the notion of {\it control kernel} within our new framework.  Therefore, our unified framework, while being suitable to mechanistic studies of GRN mutations over evolutionary timescales, is still able to describe developmental change.

The manuscript is structured as follows: The next section contains a brief review of dynamical \rv{\rv{Boolean}} models of GRNs sufficient to orient those not familiar with the general theory. It reviews the main features of Kauffman's seminal theory, and of the identification between cell types and attractors. We then expose our core idea that the cell types are more properly described by collections of attractors, i.e. by macrostates of the dynamics. We then conclude the section by drawing an example network that we will use in the sections 3 and 4 as a toy model to explore the consequences of our hypothesis. Section 3 is devoted to the main consequence of our approach. We show how gene duplication and mutation events alter the nature of the cell types expressed as a consequence of selective pressure induced by a shifting environment. \rv{In particular, we show the connection between the presence of mutations and the open-ended evolution of the selected phenotype.} Section 4 shows how our generalized approach does not inhibit the possibility of adopting network controllability methods to describe epigenetically induced developmental change. We again use the same toy model to illustrate key concepts, and show extracellular control is now more easily achieved than in Kauffman's original framework. 

\section{Boolean Models}
This section describes the mathematical details of the \rv{Boolean}, dynamical model we will assume in the rest of this manuscript.
In living tissues, the intrinsic patterns of gene expression coupled with signaling input dictates cell fate. Both of these processes can readily be modeled by a \rv{\rv{Boolean}} network. \rv{Boolean} networks were originally proposed by Kauffman \cite{Kauffman1969} as a viable mathematical model of GRNs. They permit exploration of the complex steady-state dynamics of GRNs, where the attractors of the dynamics can be identified with different cell types/fates, e.g. quiescence, proliferation, apoptosis, differentiation, etc.

\rv{In this manuscript, we anchor our discussion to Boolean networks as they represent the simplest mathematical model exhibiting biological
and systemic properties of real GRNs
\cite{Vladimir2005,Faure2006}. By virtue of this simplicity they are particularly easy to interpret biologically.}

\subsection{Review of Boolean Models}

At a given time $t$, the state of the \rv{Boolean} model of a GRN is known when the state $x_i(t)$ of every gene $i$ is known.
The \rv{\rv{Boolean}} nature of the model assumes two possible states for gene $i$: active, which corresponds to $x_i(t)=1$, or inactive, with $x_i(t)=0$. For a network describing the interactions among $n$ genes, the state at time $t$ is specified by an $n-$dimensional \rv{Boolean} array $x_1(t),\dots,x_n(t)$.
The dynamics of the GRN is then described by $n$ \rv{Boolean} functions $f_1,\dots,f_n$ which provide the state of the network at time $t+1$, given its state at time $t$ (synchronous update, but asynchronous models are also possible \cite{Garg2008}):
\begin{eqnarray} 
x_1 (t+1)  & = & f_1 (x_1(t),\dots,x_n(t)) \nonumber \\
\cdots \ \ \qquad &   &                  \nonumber \\
x_n  (t+1) & = & f_n (x_1(t),\dots,x_n(t)) \ , \label{system1} 
\end{eqnarray}

For $n$ genes, $2^n$ possible states (gene expression patterns) exist, and the dynamics of the network is represented by a trajectory (a time series) in the discrete space containing the totality of these states. For deterministic functions $f_i$, and because of the finite size of this state space, these trajectories will eventually converge to either a fixed state or a cycle of states. These special activation patterns take the name of {\it attractors} of the \rv{Boolean} network, and were identified by Kauffman as corresponding to the stable phenotypes of the GRN. 

Nothing has been said so far about the functional form of the \rv{Boolean}  functions $f_i$, and readers interested in an insightful analysis of how constrained such a specification is from experimental data are directed to \cite{Henry2013,Zhou2016}. For reasons that will become clearer in section 3, and given the conceptual nature of this manuscript, we will restrict ourselves to the case of \rv{\rv{Boolean}} network with thresholds:
\begin{equation}
x_i(t+1) = \textrm{sgn} \left( 
\sum_{j=1}^n a_{ji} x_j(t) - b_i
\right) \ ,
\label{updating_rule}
\end{equation}
where $a_{ji}$ is the relative weight of the regulatory signal from gene $j$ to gene $i$ (activation when $a_{ji}$ is positive, inhibition otherwise),  $b_i$ is the activation threshold of gene $i$, and sgn($x$) is a unitary step function, defined by $\textrm{sgn}(x)=0$ if $x \leq 0$ but $\textrm{sgn}(x)=1$ if $x>0$. 
In a more compact notation, we can write 
\[
X(t+1)=\textrm{sgn}
\big(A^T \cdot X(t)-B\big) \ ,
\]
where we have introduced the {\it adjacency matrix} $A=(a_{ij})$, and the columns $B=(b_i)$ and $X=(x_i)$. A useful feature of these models is that they convey the topology of the network, implicit in the definition of generic \rv{Boolean} functions $f_i$, in a very explicit way, with the edges of the network representing non-null entries of the adjacency matrix.

Now that we have laid the groundwork for the numerical models we will be using in the rest of this manuscript, let us go back to the interpretation of the attractors of a \rv{\rv{Boolean}} network as the phenotypes of a GRN. Given this identification, robustness and evolvability of GRNs can be mapped directly to the \rv{evolutionary changes} of the attractor landscape of the corresponding \rv{Boolean} model \cite{Aldana2007,Ciliberti2007a,Ciliberti2007b}. In this framework, the emergence of new phenotypes has a precise mathematical meaning as the acquisition of new attractors, which can arise due to mutations in the network structure. 

Many steady-state attractors are permitted in a \rv{Boolean} network, making it an ideal model for describing how the genomic information contained within a single initial fertilized egg is differentially expressed in so many distinct cell types in response to different regional specification and morphogenetic histories \cite{Boveri1906}. 

In the absence of external regulation, the likelihood of a given cell type, or phenotype, is quantified in terms of the number of initial configurations of the network converging on the attractor state encoding that cell type. The higher the number of initial configurations leading to the same equilibrium dynamics, i.e. the production of a well defined set of proteins, the more likely the cell type that set of proteins represents will be. We will refer to these probabilities as the {\it basins of attraction} of the possible cell types encoded in the GRN.

On the other hand, the effect that external regulation, in the form of extracellular signaling during development, might have in determining the expression of a specific subset of genes, is even more dramatic, as it might force the development toward cell types that were not even initially accessible to the unperturbed GRN. The minimum number of genes whose expression needs to be controlled, for the cell fate to be determined, is the {\it control kernel} associated to that cell type \cite{Kim2013}, as described in the introduction.   

\subsection{GRNs as Evolvable Systems}

Most of the predictions derived within the framework we have just reviewed rely on the assumption that different cell phenotypes correspond to different attractors of the GRN. While the connection between cell types and attractors is both numerically and experimentally well-motivated, their one-to-one correspondence might be an \rv{artifact} due to the diminutive size of the mathematical models that were actually solvable in the not-so-distant past, when a small number of attractors were naturally identified with different phenotypes. Recent advances in computing power are finally making the study of much larger networks possible. One interesting example is the recent dynamical model developed by Fumi\~a and Martins \cite{Fumia} for the integration of the main signaling pathways involved in cancer. The signaling among almost 100 different genes is responsible for 63 different attractors, that eventually correspond to {\it only three} phenotypically distinct and incompatible cell fates: apoptotic, quiescent, and proliferative. An important observation is that these phenotypes are determined by just a small subset of values, e.g. the constant activation of the effector caspases in apoptotic cells, within a much larger gene activation pattern.

A second example of several attractors all sharing the same phenotypic identity is already present in the much smaller GRN describing cell-fate determination during Arabidopsis thaliana flower development \cite{Espinosa-Soto2004}. Four among ten possible attractors all represent the same inflorescence meristematic cell type.

Therefore, while most of the literature on network robustness and evolvability focuses on individual attractors, the concept of phenotype --that we redefine here as a macro-state of the cell characterized by the activation values (fixed or cycling) of only a subset of gene nodes-- seems to be better suited to the idealization of biological systems. While an attractor can represent a phenotype by itself, a phenotype can be compatible with multiple attractors. 

Our definition is exemplified in the following diagram, showing the case of a phenotype $\Phi$ defined by the expression of the two genes represented by dark gray boxes, and the \rv{non expression} of the two genes shown in light gray tone:
\begin{center}
\includegraphics[width=.9\columnwidth]{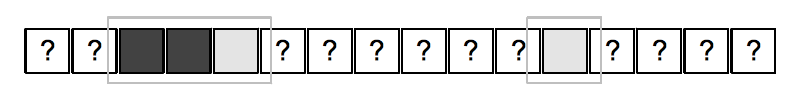}
\end{center}
The expression of any other gene (boxes with a question mark) can take any value, as long as the four genes entering the definition of $\Phi$ exhibit the right expression pattern.\\

\begin{figure*}[!t]
\centering
\includegraphics[height=.55\textwidth]{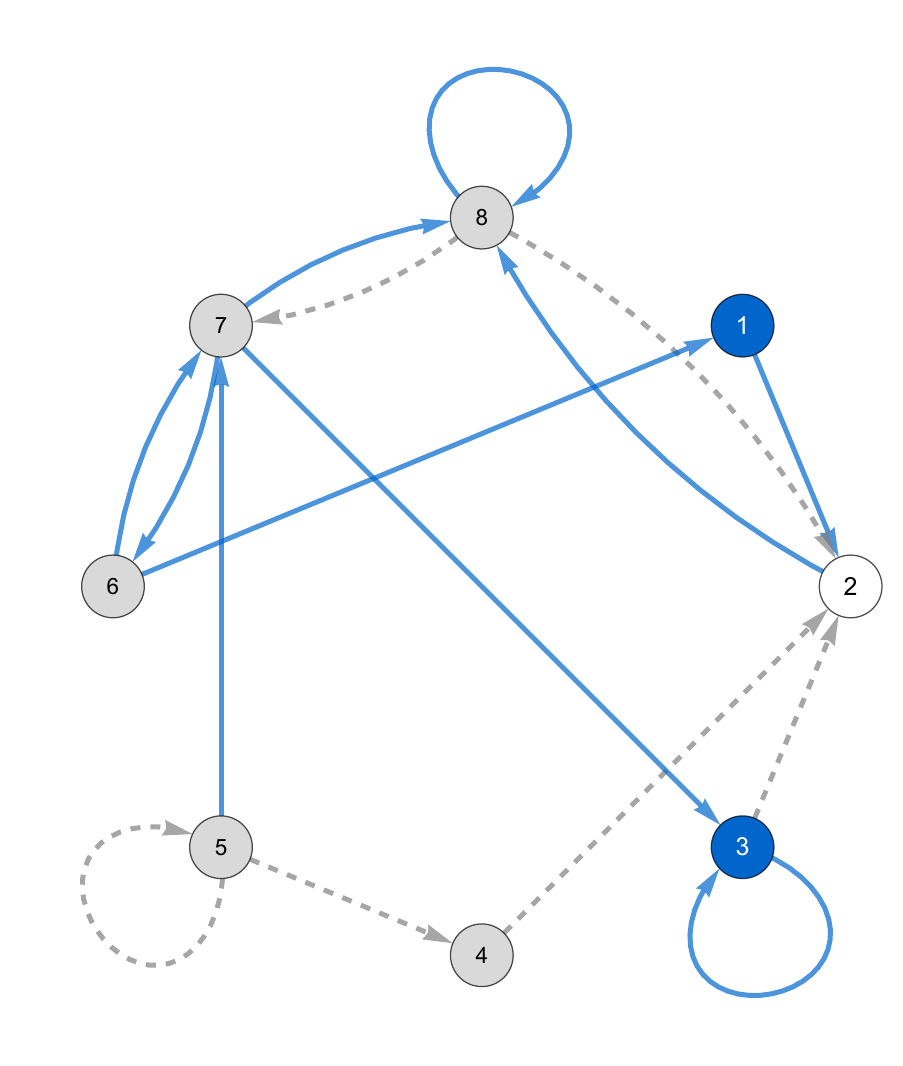}
\caption{\rv{Boolean threshold network (as described in subsection 2.1) for the GRN of cell type 1. The subset of genes defining the phenotype are represented by nodes 1, 2, and 3. The characteristic pattern of cell type 1 is assumed to have genes 1 and 3 expressed, and inactive gene 2 (blue = active, white = inactive). In order to reproduce our example, the reader needs to assume equal weight edges $a_{ij}=\pm 1$ for activating/inhibiting links (continuous/dashed line), non-zero thresholds $b_1=-1$ and $b_{5,6,8}=1$.}}
\label{GRN}
\end{figure*}

A critical motivation for this redefinition of the mathematical identity of a phenotype is a conceptual difficulty which naturally arises in Kauffman's original framework \cite{Kauffman1969}. Kauffman's definition loses its strict meaning in light of evolutionary biology: the identification of the attractors of a \rv{\rv{Boolean}} network with the phenotypes of a GRN requires --after the regulatory network has lost or acquired genes-- the comparison between genotypes of different lengths.

An immediate, advantageous consequence of our definition is it keeps its meaning even after changes occur in the genotype. As an example, the following diagram, shows the case of a single gene duplication event enlarging the genome expressing 
$\Phi$:
\begin{center}
\includegraphics[width=.9\columnwidth]{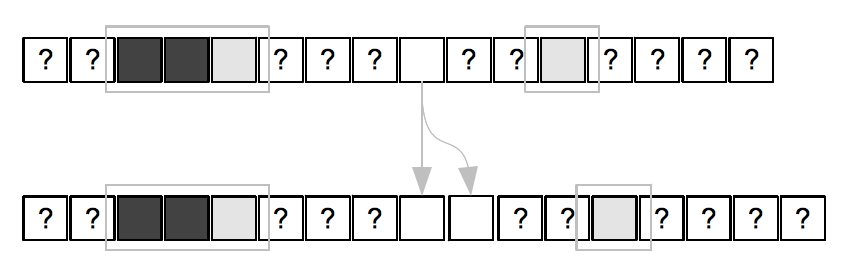}
\end{center}
As this mutation is not affecting any of the defining genes, this change in the size of the genome does not alter our definition of $\Phi$. Therefore, it makes sense to study whether $\Phi$ is still expressed by the new GRN, and whether the gene replica is favoring or disfavoring $\Phi$'s basin. As we will show in section 3, these questions can be addressed in quantitative terms. The inclusion of different gene replicas will induce different changes in the relative size of the basin of attraction of $\Phi$, and a good dynamical model of the GRN will be enough to determine the duplication events that increase the basin of $\Phi$. When $\Phi$ is selected, these are likely to be the duplication events that are fixed more often. 

A second, less relevant difference between our approach and those outlined in previous literature on GRNs, is that we will not require the basin of attraction of a specific cell type to be exactly 100\% for the network to be representative of that specific type. Instead, we will only require one basin to be much larger than the remaining others. In the following example, aimed at making these ideas more concrete, we will assume 80\% as the threshold a basin needs to exceed for the remaining attractors to be treated as negligible.

\subsection{A Sample Network}

In the next two sections, we will explore the consequences of our approach in studying evolutionary and developmental processes \rv{within} the \rv{\rv{Boolean}} model framework. To anchor our discussion to a concrete example, we will introduce a small, tractable network that we will adopt as a toy model, and modify as needed.

The example is shown in figure \ref{GRN}. It represents the GRN expressing a phenotype that we will call {\it cell type} 1. The subset of genes defining the phenotype are represented by nodes 1, 2, and 3. The characteristic pattern of cell type 1 is assumed to have genes 1 and 3 expressed, and inactive gene 2 (blue = active, white = inactive in figure \ref{GRN}). We will refer to this by saying that cell type 1 has expression pattern 101. In determining the basin of attraction of cell type 1, we will sum the basins of every attractor of the GRN which is compatible with the expression pattern 101, i.e. every pattern of the form $1,0,1,x_4,\dots, x_8$, where $x_4,\cdots,x_8$ (gray nodes) can take any possible (binary) value representing the expression/suppression of the remaining genes labeled from 4 to 8.

The GRN of figure \ref{GRN} actually encodes two possible cell types, one being type 1, as we said, the other having the activation pattern 110. We will refer to this second phenotype as {\it cell type} 2. Despite encoding both phenotypes, the basin of attraction of cell type 1 includes more than 80\% of the possible initial activation patterns, all leading to the equilibrium dynamics characterized by the expression of genes 1 and 3, and the suppression of gene 2. This is why we assume this network to be a viable description of cell type 1. 

For the reader interested in reproducing our example: The GRN dynamics is modeled using a \rv{\rv{Boolean}} threshold network (as described in section 2.1), with equal weight edges $a_{ij}=\pm 1$ for activating/inhibiting links (continuous/dashed line in figure \ref{GRN}), non-zero thresholds $b_1=-1$ and $b_{5,6,8}=1$. But it is important to remark that none of the conclusions we will draw in the next sections depend on the mathematical details of the example we are using.

\section{Evolution}

We have seen how our re-definition of the phenotypes of a GRN in the framework of \rv{Boolean}, dynamical models is \rv{relatively robust towards mutations changing the size of the genome}. In this section, we seek to show an explicit example of how this allows the prediction of the genes whose duplication will reinforce a phenotype favored by natural selection. The underlying idea is exemplified in the following diagram (same example of phenotype $\Phi$ adopted in subsection 2.2), with arrows showing the regulation induced by an ideally connected \rv{gene} which activates exactly the \rv{genes} that are supposedly  expressed, and suppresses just the genes that are inactive in our former definition of $\Phi$: 
\begin{center}
\includegraphics[width=.9\columnwidth]{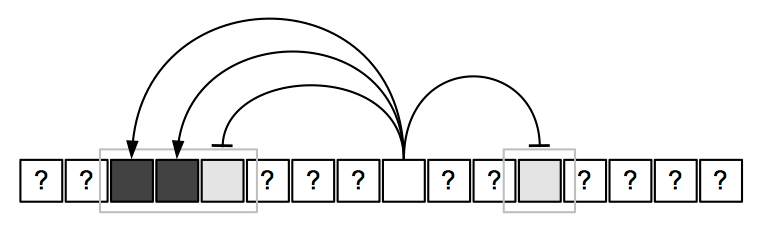}
\end{center}
It is easy to foresee that the duplication of this ideal \rv{gene} would positively impact the change in the basin of attraction of $\Phi$. In a more realistic scenario, duplication would be followed by divergence, represented in the following diagram by a difference in one of the genes regulated by the replica: 
\begin{center}
\includegraphics[width=.9\columnwidth]{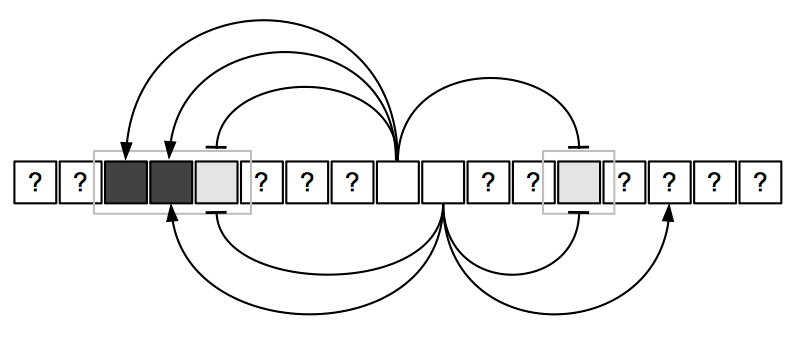}
\end{center}

\begin{figure*}[!t]
\centering
\includegraphics[width=.6\textwidth]{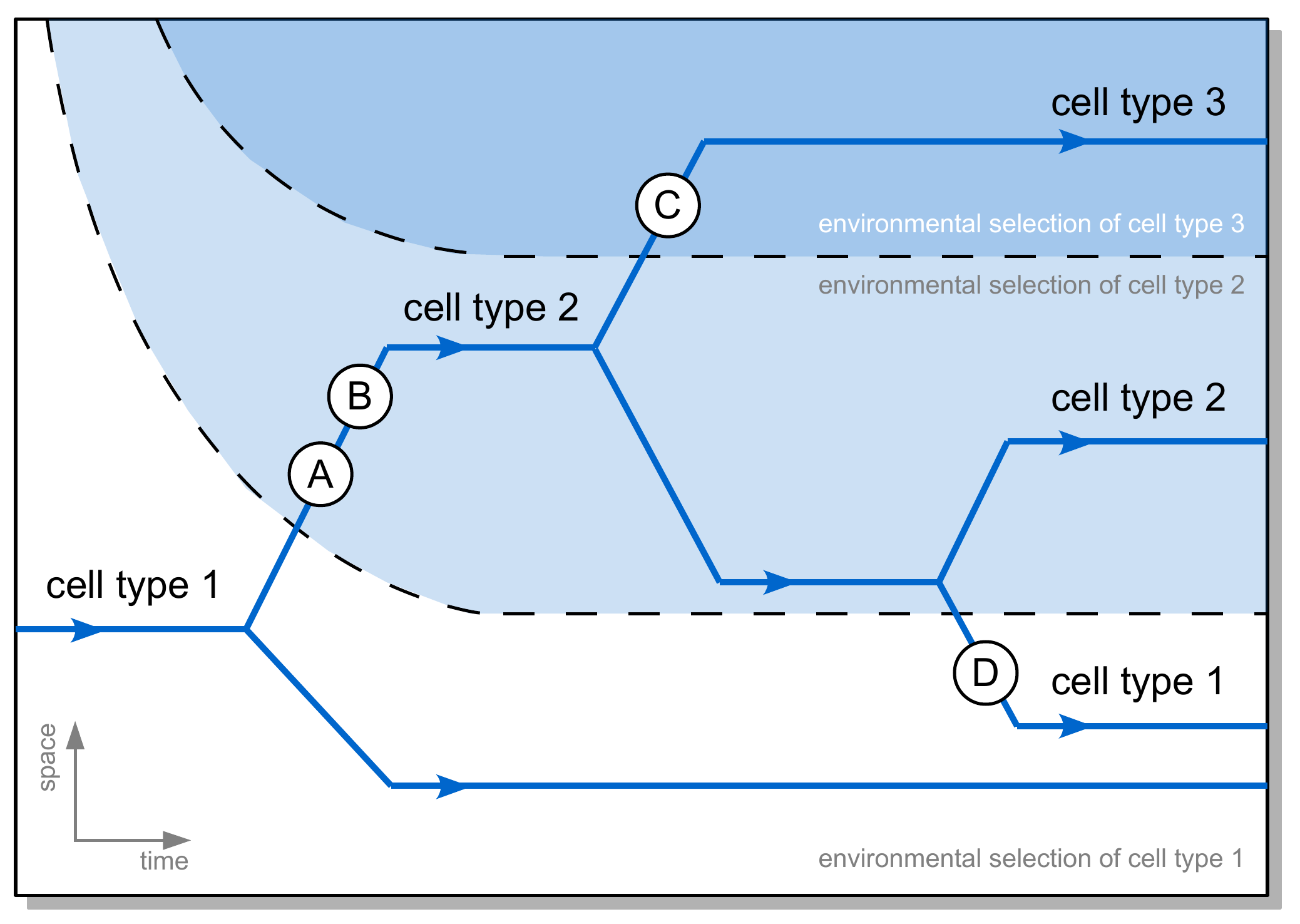}
\caption{Different environmental pressure in both space and time selects cells undergoing mutation events that favor the emergence of newly preferred cell types.}
\label{selection}
\end{figure*}

In this section we will consider duplication events like the one just described. For each \rv{gene that does not enter the definition of the phenotypes} in the GRN, we will consider the entire spectrum of possible events of {\it duplication} + {\it divergence}, and show ideal pathways leading to cell type changes in a shifting environment.
With reference to figure \ref{selection}, we will describe a simple evolutionary model for the differentiation of progenitor cells of type 1, as encoded in our network from subsection 2.3, into cells of type 2 (already encoded, but unlikely), and the newly born {\it cell type} 3, a novel phenotype induced by the modifications the network is going through. 

We are assuming here that the knowledge of the environment is equivalent to the knowledge of the selected cell type: they are on the same ontological level, and {\it share the same mathematical definition} in our theoretical model. The environmental pressure changes in both space and time, and selects cells that undergo mutation events that favor the emergence of the newly preferred cell type. In this example, sister cells \cite{Arendt2008} of type 1 start being selected for cell type 2, and undergo mutation events \circled{A} and \circled{B} until cell type 2 represents the dominant phenotype. In our model, the mutation events are represented by gene duplication and divergence of one of the \rv{genes not included in the definition of the phenotypes} (nodes 4--8 in fig. \ref{GRN}), during which the network acquires a non-mutated replica of a preexistent \rv{gene}, and then a mutation in the way the replica regulates the remaining genes and itself. 

Let us first provide a mathematical description of what we mean by a perfect replica of a preexistent gene. We will then introduce divergence in the form of a mutation affecting the replica's downstream signaling. \rv{Lastly, we will show that open-ended evolution is only possible if the network can acquire mutated gene replicas. This digression well highlight the general features of our model, before we show them in action on our specific example.} \\

\noindent
{\bf Non-mutated gene replica\rv{s}:} Given a \rv{\rv{Boolean}} network $N$ with $n$ nodes, we want to study the dynamics of the $n+1$-node network $N'$ obtained by the inclusion of the perfect replica of a node already present in $N$. 

If the dynamics of $N$ is governed by the set of \rv{\rv{Boolean}} equations 
\begin{eqnarray} 
x'_1  & = & f_1 (x_1,\dots,x_n) \nonumber \\
\dots &   &                  \nonumber \\
x'_n  & = & f_n (x_1,\dots,x_n) \ , \label{system1} 
\end{eqnarray}
then the updating rules of $N'$ will have the form
\begin{eqnarray} 
x'_1  & = & \rv{g}_1 (x_1,\dots,x_n,x_{n+1}\rv{)} \nonumber \\
\dots &   &                  \nonumber \\
x'_n  & = & \rv{g}_n (x_1,\dots,x_n,x_{n+1}\rv{)} \nonumber  \\ 
x'_{n+1}  & = & \rv{g}_{n+1} (x_1,\dots,x_n,x_{n+1}\rv{)} \nonumber
\label{system2} 
\end{eqnarray}
In the previous equations, $x_i$ is the value (0 or 1) of node $i$ at time $t$, while $x'_i$ is the simplified notation for the value of the same node at time $t+1$. 
The $\rv{g}_i$ functions are {\it new} functions that we want to determine, given our knowledge of the functions $f_i$. Without loss of generality, we can assume node $n+1$ to be a perfect replica of node 1.
Let us then consider the requirements we impose on $\rv{g}_i(x_1,\dots,x_n,x_{n+1})$, after we include node $n+1$:

\begin{enumerate}

\item The assertion that node ${n+1}$ is a {\it perfect replica} of node $1$ translates into the assumption that 
\[
\ \rv{g}_{n+1}=\rv{g}_1.
\] 
Here we are just stating that node replicas obey the same updating rules as the original nodes.

\item We also want the new node not to affect the system, when not expressed. Therefore,
\[ 
\rv{g}_i (x_1,\dots, x_n, 0) = f_i (x_1,\dots, x_n) \qquad  (i=1,\dots,n)
\]

\item Lastly, we want nodes $2,\dots,n$ to see node 1 and $n+1$ as {\it indistinguishable}:
\begin{eqnarray}
\rv{g}_i (0,x_2,\dots, x_n, x_{n+1}) = f_i  (x_{n+1},x_2,\dots, x_n) \nonumber \\ (i=1,\dots,n) \nonumber
\end{eqnarray}

\end{enumerate}

The previous conditions are enough to determine the values taken by the functions $\rv{g}_i$ for any dynamical state of $N'$ with the only exception being those cases where $x_1$ and $x_{n+1}$ are both 1. This is the only genuinely new scenario whose output cannot be predicted in terms of an equivalent configuration of $N$. The problem is easily solved in the case of \rv{\rv{Boolean}} networks with thresholds (subsection 2.1), as we know the prescription that gives the updating rule of a node in terms of the state of the network, eq. (\ref{updating_rule}).  Building the functions $\rv{g}_i$ in a similar fashion, we deduce that it is both natural, and enough, to impose $a_{j+n,i}=a_{ji}$ and $b_{i+n}=b_i$ for the previous three conditions to be satisfied. \rv{This prescription, first adopted in \cite{Wagner1994}, is re-derive here, and contextualized into a more general framework of GRN described by arbitrary \rv{\rv{Boolean}} functions.} \\

\begin{figure*}[t]
\centering
\includegraphics[width=.5\textwidth]{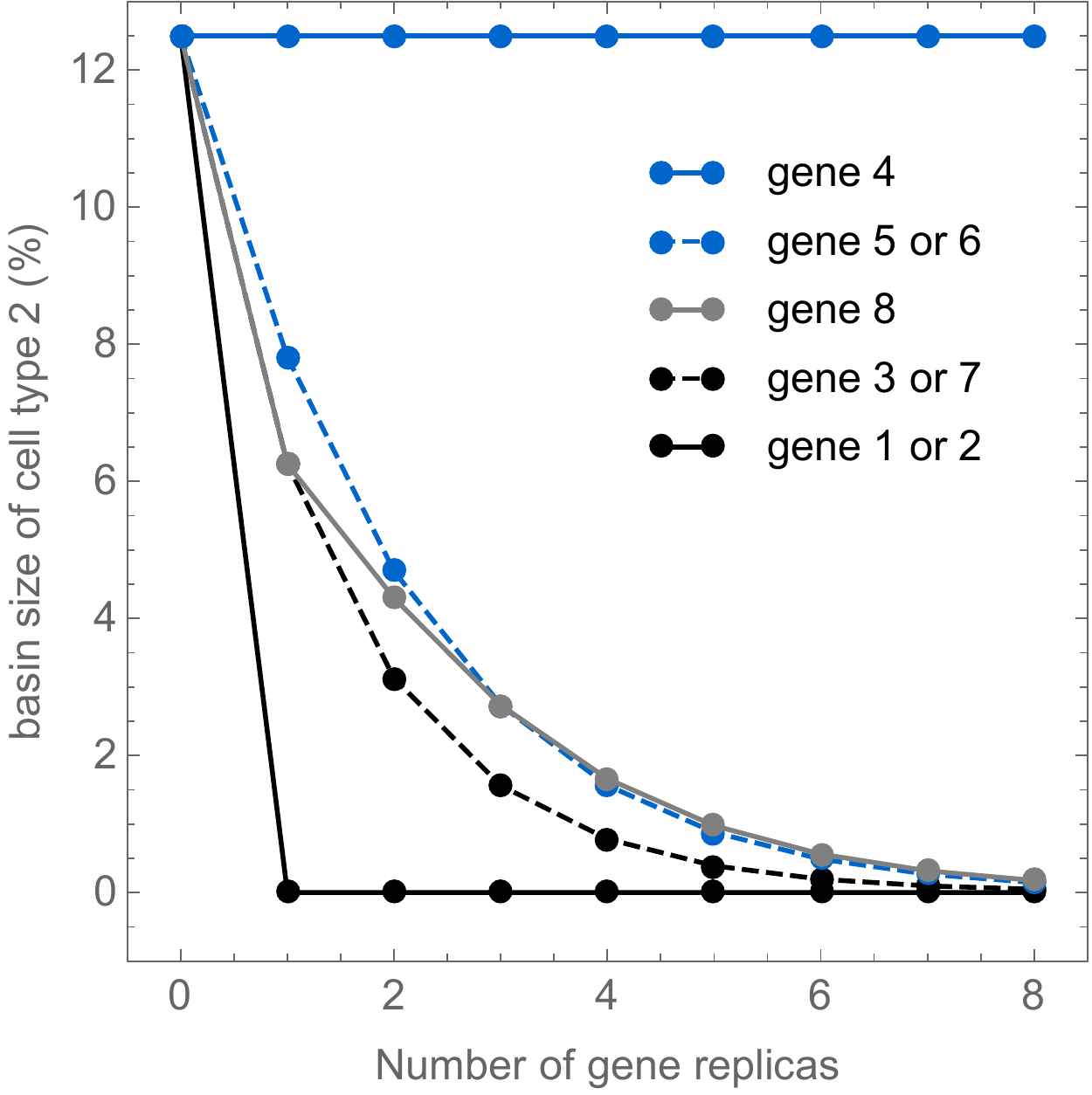}
\caption{\rv{Effect on the basin size of cell type 2 of multiple non-mutated gene replicas.}}
\label{mul_repl}
\end{figure*}

\noindent
{\bf Mutation\rv{s}:} The signaling of gene replica $i$ is represented by the string of numbers $a_{ij}=0,{\pm}1$, with $j=1,\dots,n$ ($n$~\,=~\,current number of genes in the network). A mutation is introduced by changing one of these numbers to \rv{any one} possible value \rv{(including its initial one, which corresponds to having a perfect replica)}. For example, changing $a_{ij}$ from 0 to $\pm1$ corresponds to the creation of a new edge. Changing it from $\pm1$ to 0 means deleting an edge which is instead departing from the original node. Changing it from $\pm1$ to $\mp1$ changes an activating link into an inhibiting one or vice-versa. Different models of mutation could be assumed, which in turn determine the minimum numbers of replicas the cell needs to acquire in order to change from one type to another. For simplicity, our examples show only the sequence of mutations that minimizes the number of steps needed to complete the phenotypic shift. Therefore, we consider all possible mutations of the kind previously described, and then select the one that maximizes the basin of attraction of the preferred cell type.\\

\rv{

\noindent
{\bf Mutations and open-ended evolution:}
In this section we want to show that allowing divergence (mutations of gene replicas) guarantees open-ended evolution of the GRN. While there are many definitions for open-ended evolution, here we restrict the concept to a simple one, and define a GRN as ``open-ended'' if the GRN can acquire any possible phenotype, given a high enough number of mutated gene replicas (this becomes open-ended in the sense the mechanism is robust to any evolutionary innovation). On the contrary, we will show that, if mutations were not possible, exact gene replicas would be unable to induce arbitrary phenotypic shifts.

Let us begin by showing that mutations enable open-ended evolution. At the beginning of this section we have referred to the possibility that the GRN acquires replicas of genes optimally targeting the nodes entering the definition of a specific phenotype, and, therefore, reinforces it. Here we want to show a simple conceptual procedure able to create such optimal genes. We will focus on the existence of a path toward the expression of a novel phenotype, not on the optimal/fastest evolutionary path that leads to that. Such optimal path will depend on the specific network under study. And we will show with other example that it can be much shorter that the general construction illustrated here.

Let us label $i_1, \dots, i_p$ the set of $p$ genes that specify the {\it arbitrary} phenotype $\Phi$. For example gene $i_1$ might need to be constantly active, $i_2$ constantly inactive, etc. Starting from an equally {\it arbitrary} gene $j$ with $m$ outgoing edges in the GRN, we can consider $m$ subsequent events of duplication and mutation of $j$ into genes $j_1, j_2, \dots j_{m}=j'$. Each mutation is affecting one of the downstream signals produced by $j$, until $j'$ activates only itself and $i_1$. Each additional, non-mutated replica of $j'$ will therefore activate itself, $i_1$, and each other replica of itself: 

\begin{center}
\includegraphics[width=.9\columnwidth]{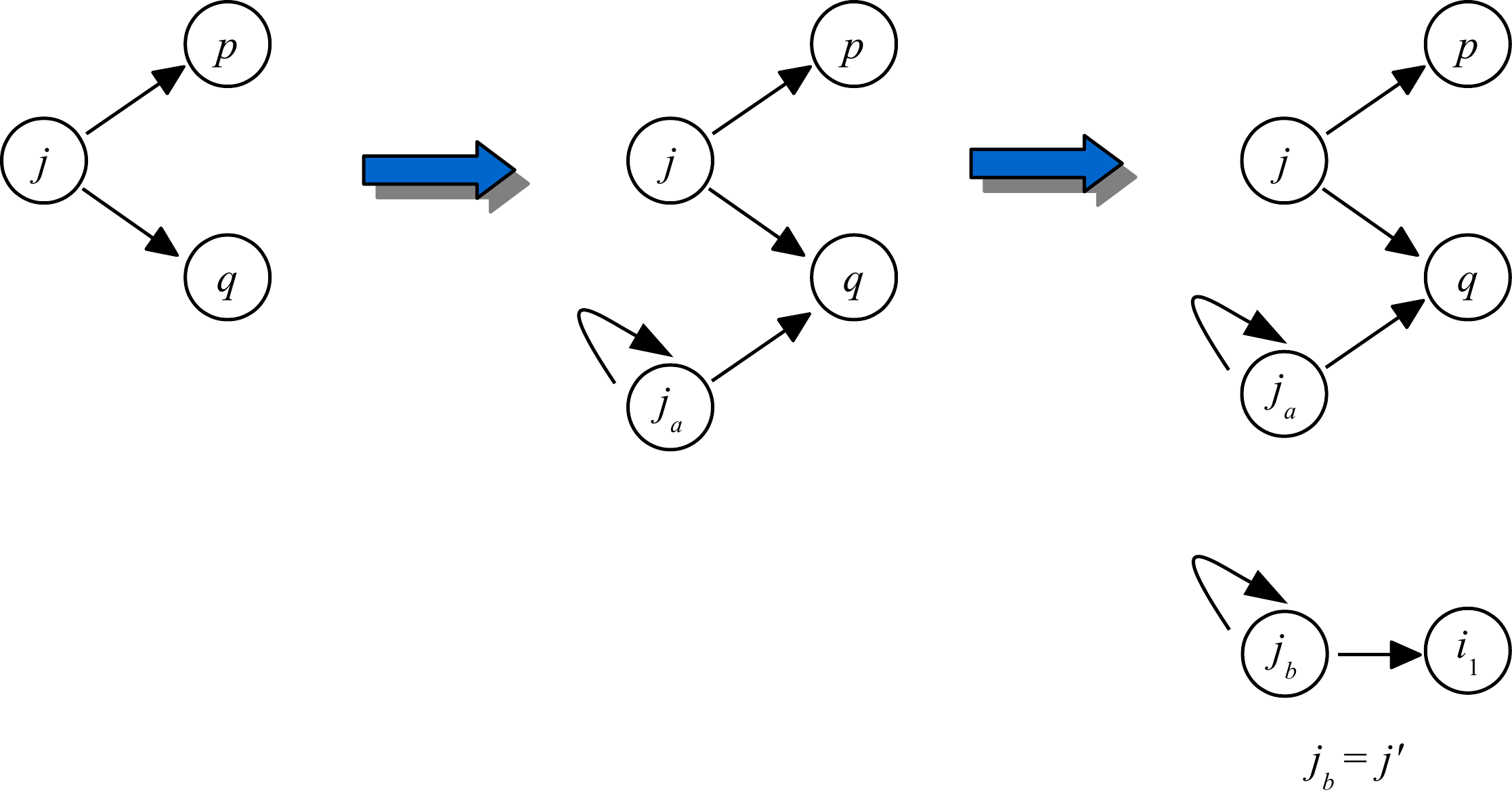}
\end{center}

At this point it is possible to construct a clique of self-activating $j'$ nodes that keeps itself constantly on, regardless of the signaling the clique receives from the remaining genes in the GRN. For this to happen it is enough to have $n_1$ copies of $j'$ with 
$n_1 > b_1 + w_1$, 
where $b_1$ is the activation threshold of $i_1$, and $w_1$ is the number of inhibiting incoming edges of $i_1$: 

\begin{center}
\includegraphics[width=.9\columnwidth]{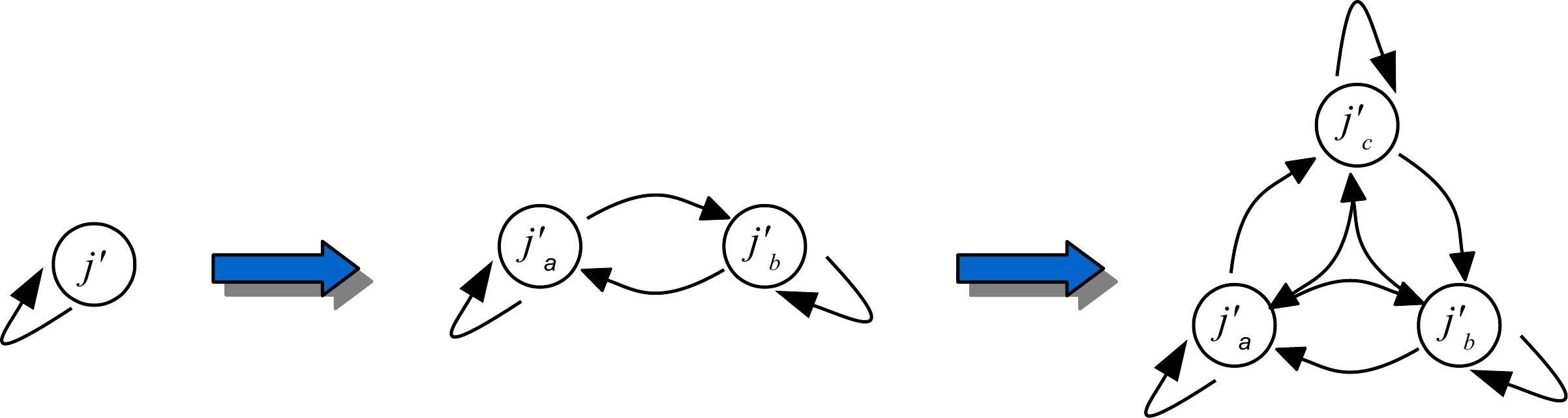}
\end{center}

A similar construction guarantees the possibility of keeping gene $i_2$ inactive by simply having $j'$ inhibiting it, as opposed to activating it.

The previous construction is intentionally artificial, as we wanted to show that a complete shift of the GRN toward and arbitrary phenotype $\Phi$ can be achieved starting from the duplication and mutation of an arbitrary gene $j$ in the network. Even in the simplest examples, this kind of construction would require many events of duplication + mutations. Our aim here is to demonstrate the existence of an evolutionary path toward $\Phi$. In our example, we consider all possible duplication + divergence events, and only show the ones that induce the fastest shift, as we now know that many other, and slower paths, toward the same phenotypic shift exist. 

As a final remark, we point out less restrictive choices of the possible mutation events, as well as the inclusion of deletion events, just increase the number of paths towards phenotype $\Phi$.\\

\noindent
We now will show open-ended evolution is not possible in the absence of mutations. To this aim, it will be enough to provide a counter-example. Starting from our reference network (figure \ref{GRN}), we consider the effect that the creation of any possible number of exact replicas would have, and we will show that it is not possible to choose an arbitrary phenotype $\Phi$, and convert it into the dominant phenotype. Let us assume $\Phi =$ cell type 2. With reference to figure \ref{mul_repl}, we can see that the relative basin size of cell type 2 does not depend on the number of additional, non-mutated replicas of gene 4. A single replica of gene 1 or 2 is instead enough to completely extinguish the phenotype. Additional replicas of the remaining nodes slowly decreases the size of the basin. Perfect replicas of any gene in the network are either negatively affect the likelihood of cell type 2, or not affect it at all. Therefore, it is not possible to induce a phenotypic shift from cell type 1 to cell type 2 by duplicating genes already present in this GRN.

Analogously, it is not possible to create an arbitrary new phenotype through non-mutated gene replicas. As an example we can consider the cell type defined by the activation pattern $100$. Let us call it {\it cell type} 3. We know that this phenotype has an empty basin of attraction, and direct calculation shows that the exact duplication of {\it any} gene in the network leaves its basin unchanged.

On the other hand, we are about to show that even adopting just our very restrictive prescription of divergence it is possible to mutate of example network until it undergoes a complete shift from cell type 1 to cell type 2, and even 3. (In agreement with the general argument just exposed.)\\
}

\noindent
\rv{Therefore, let us} return to the example shown in fig. 3 \rv{to} deduce that at least two mutation events are needed to convert sister cells of type 1 to type 2 (activation pattern: 110). 
\rv{These optimal mutations are derived as previously described in this section: All possible events of duplication + mutation are considered, and the one inducing the largest enhancement of the relative size of the basin of cell type 2 is selected. The network is then mutated accordingly, and this process iterated until cell type 2 is the dominant phenotype.}

\begin{figure*}
\centering
\begin{tabular}{|c|c|} \hline & \\
\begin{tabular}{rl} 
\circled{A} & Environmental selection of {\bf cell type 2},\\ 
& characterized by the expression of gene 1\\
& and 2, and suppression of gene 3. Acqui-\\ 
& sition of a mutated replica of gene 7 that\\
& inactivates gene 3 instead of activating\\ 
&  it. Mixed cell types (50\% type 1, 50\% \\ 
&  type 2).
\end{tabular}
&  
\begin{tabular}{rl}
\circled{B} & Environmental selection of cell type 2.\\
& Acquisition of a mutated replica of gene   \\
& 6 that does not activate gene 7. Type 2 \\
&  cells represent now the dominant cell \\
& type (88\%).\\
& \\
& 
\end{tabular}
\\
\includegraphics[height=.30\textheight]{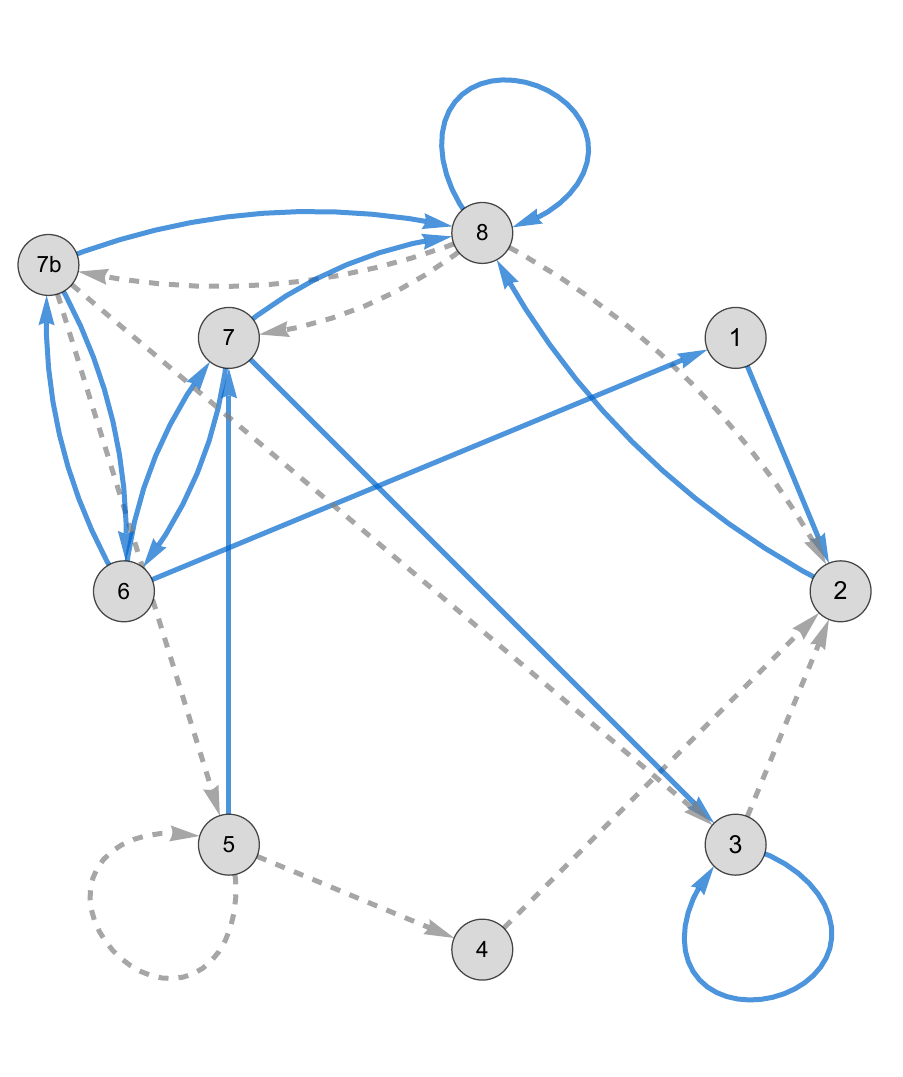} &
\includegraphics[height=.30\textheight]{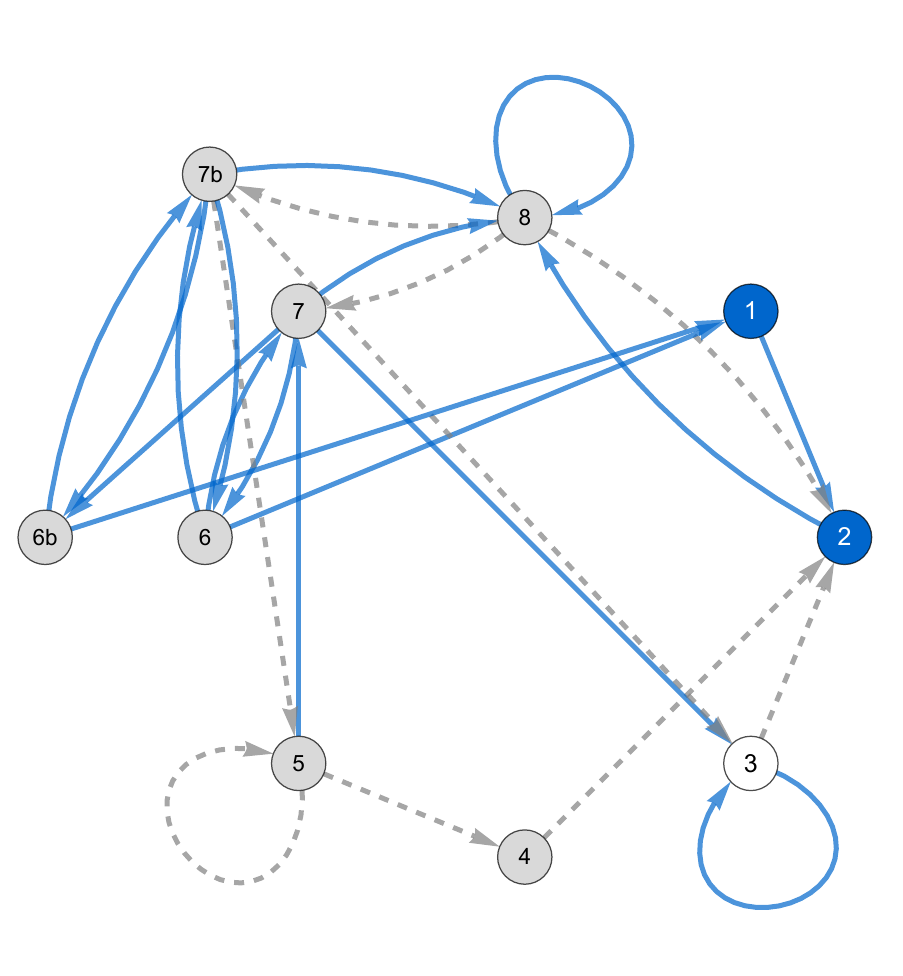} \\[2mm] \hline 
 & \\
\begin{tabular}{rl}
\circled{C} & Environmental selection of {\bf cell type 3}:\\
& Expression of gene 1, suppression of gene 2   \\
& and 3. Acquisition of a mutated replica of \\
& gene 8 that suppresses gene 3. Type 3 be-\\
& comes the dominant cell type (92\%).\\
& \\
& 
\end{tabular}
& 
\begin{tabular}{rl}
\circled{D} &  After having been selected for type 2,\\
& cells are once again selected for {\bf type 1}. \\
& Acquisition of a mutated replica of gene\\
&  8 that activates gene 3. Type 1 is again\\
& the dominant cell type (96\%).\\
& \\
&
\end{tabular}
\\
\includegraphics[height=.34\textheight]{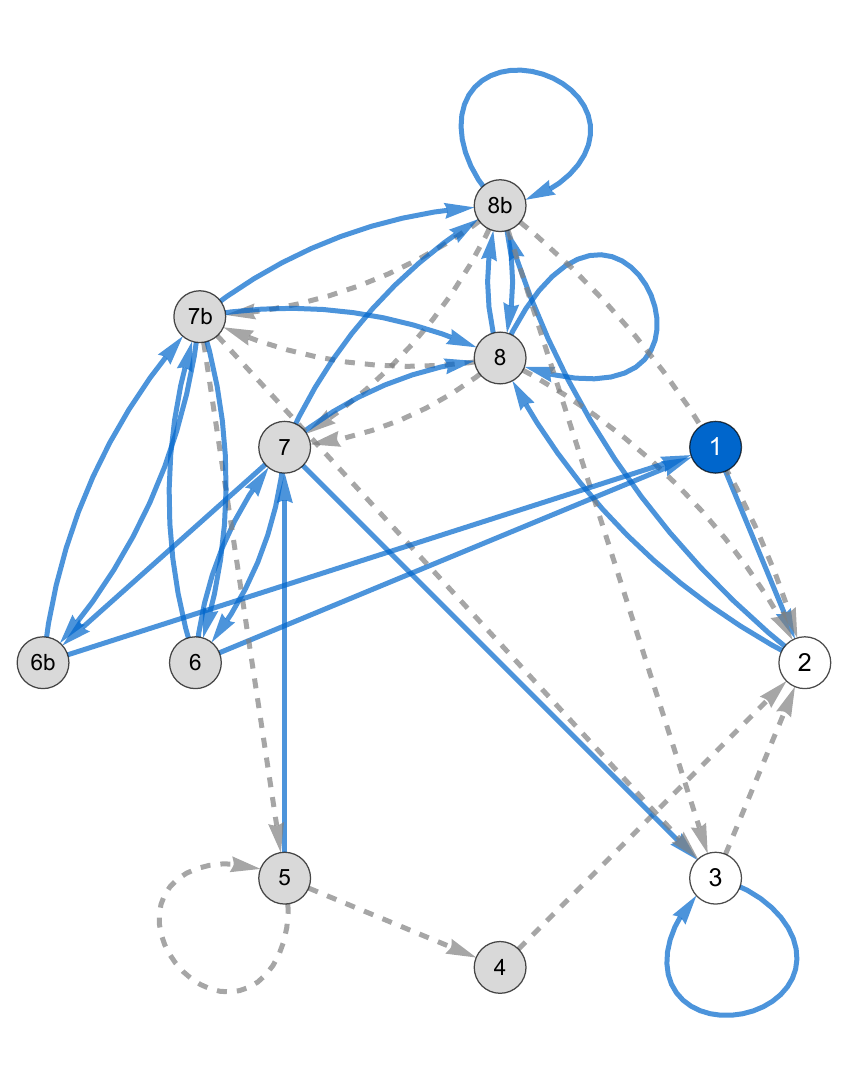} &
\includegraphics[height=.34\textheight]{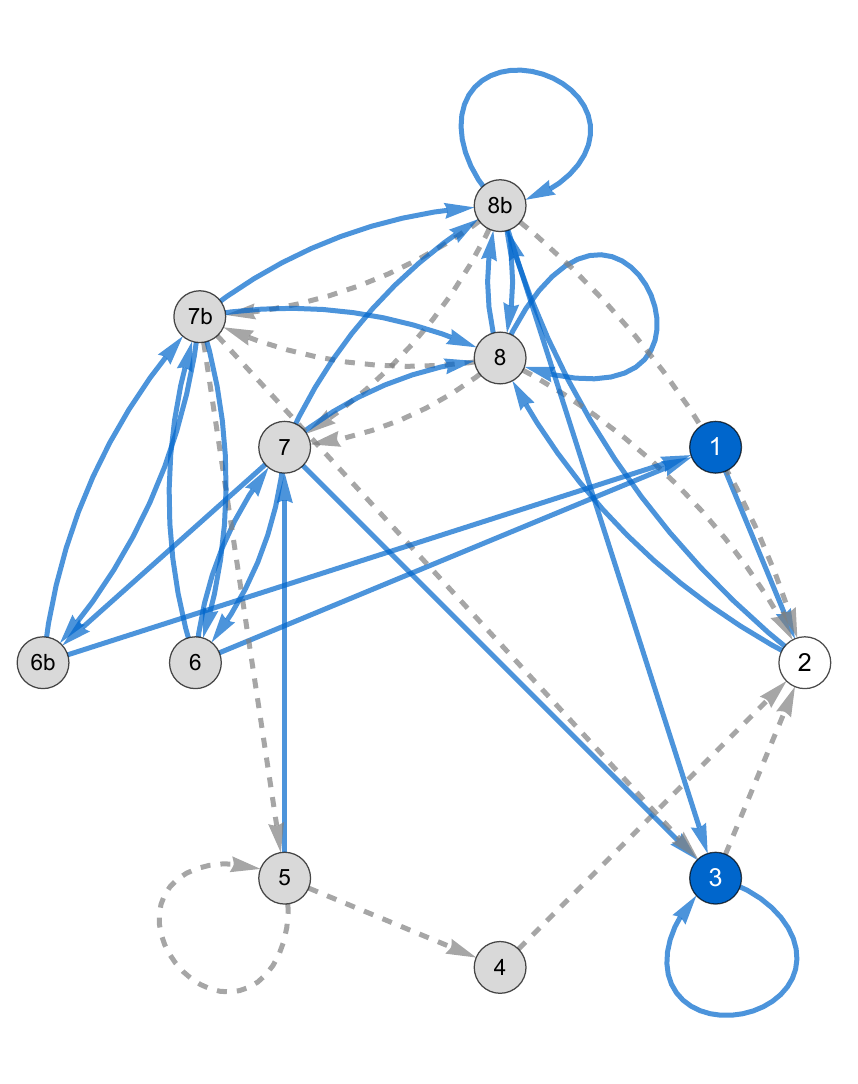} \\ 
& \\ \hline
\end{tabular}
\caption{Duplication and mutation events leading from cell type 1 to 2, and 3, and from cell type 2 back to 1, as described in section 3.}
\label{ABCD}
\end{figure*}

By acquiring a mutated replica of \rv{gene} 7, event \circled{A}, that inactivates gene 3 instead of activating it, the dynamics of the GRN is equally likely to converge toward cell type 1 or 2 (figure \ref{ABCD}). 
The subsequent acquisition of a mutated replica of gene 6, event \circled{B}, that, differently from the original one, does not activate gene 7, determines the complete shift toward cell type 2 (88\% of the configuration space). 
The highly connected subset of nodes that appear in the B-panel of figure \ref{ABCD} is responsible for the convergence of the dynamics toward the characteristic activation pattern which defines cell type 2.

The assumption that sister cells of cell type 2 start being selected for  cell type 3 (activation pattern: 100), determines the selection of mutation events like \circled{C}, where the GRN acquires a mutated replica of \rv{gene} 8 that suppresses gene 3. One mutation event is now enough to determine the complete shift toward cell type 3. 

The last example we will consider here is the convergence of cell type 2 back to cell type 1. After the acquisition of mutations \circled{A} and \circled{B}, the preferred cell type is again type 1. This can be achieved (complete shift from less than 12\% to more than 96\%) with a single mutation event, marked \circled{D} in figure 3, consisting of a mutated replica of \rv{gene} 8 that activates gene 3. This last evolution of the GRN expresses the same phenotype as the initial network in figure \ref{GRN}, but carries memory of the evolutionary path that led it through its  type 2 period in the form of the \rv{modular structure} visible in the D-panel of figure \ref{ABCD}.

We hope this example is enough to convince the reader that, as long as a viable mathematical description of the GRN is available, as well as the knowledge of the preferred phenotypes for a shifting environment, our approach \rv{can identify the events of gene duplication and mutation with the highest causal effect on the phenotipic shift.} The possibility to retain cell type mathematical identities for evolving GRNs is the key feature enabling this. 

We have also shown that our method allows the identification of GRNs that differ for both genotype and network topology (like the GRN in figure 1 and the evolved network in the D-panel of figure 3) as encoding {\it the same cell type} (type 1 in our example). Therefore, apparently redundant topological features in the structure of a GRN might carry the imprint of their evolutionary history, and of the environment that induced them.

\section{Development} 

\begin{table*}[t]
\centering
\includegraphics[width=.5\textwidth]{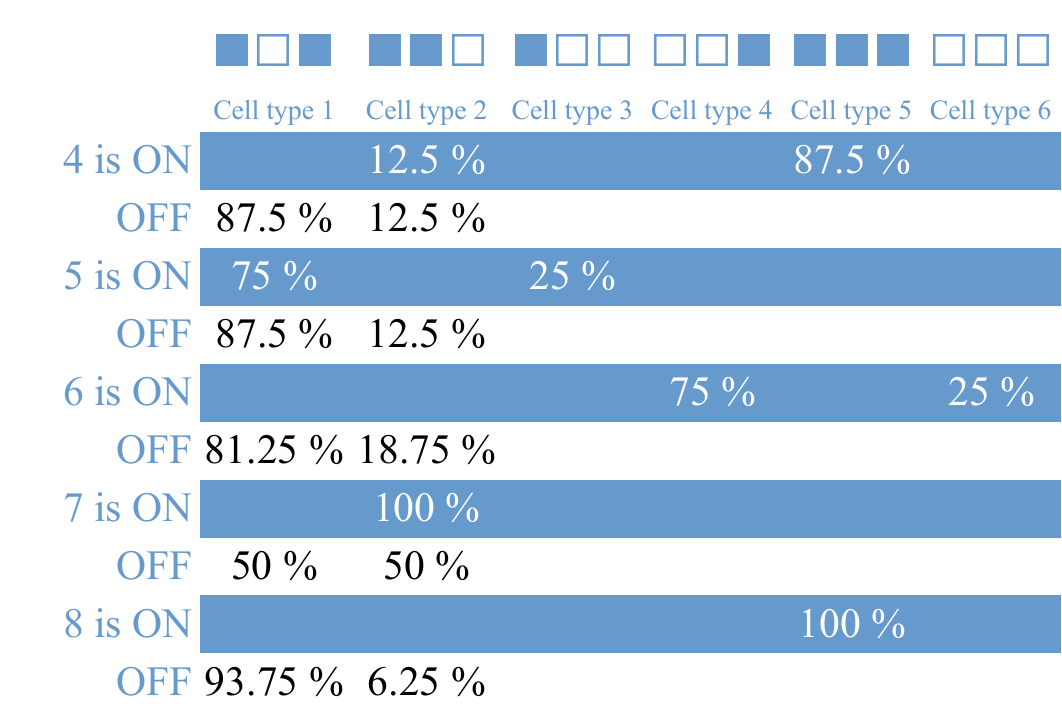}
\label{pinning1}
\caption{Phenotypic change induced by the controlled expression of \rv{genes 4-7}. This table shows the result of pinning just one \rv{gene} at a time (TS ON/OFF). Four new phenotypes, cell types $3\div6$, not initially encoded in the GRN of figure \ref{GRN}, are now possible equilibrium state of the gene expression dynamics, with basins of attraction sizes (likelihoods) listed as percentages.}
\end{table*}

Next we want to show how external control of the expression of accurately chosen genes -- e.g. as under the effect of drug therapies, or because of extracellular signalling during development-- can determine cell differentiation, and the expression of not only cell type 2, but also novel cell types that were not initially encoded in our sample network.

For an explanation of cell differentiation not invoking extracellular signaling, and relying instead on cellular noise, the reader is redirected to \cite{Villani2011}.
We will focus here on the possibility of epigenetic signaling to induce the appearance of new cell types. This approach has been extensively investigated as an alternative explanation \cite{Huang2012,Pisco2013} to the current paradigm of treatment-selected, drug-resistant clones in tumor progression \cite{Attolini2009,Diaz2012}.
We generalize it here to the broader problem of cell differentiation  over a developmental timescale. We also show the role played by our generalized definition of phenotype in reducing the size of the {\it control kernel}, as defined in \cite{Kim2013}. 

The original approach developed in \cite{Kim2013}, consists of taking every possible combination of any subset of genes and their possible expression patterns. For each combination, and each initial state, time series of the transient expression pattern of the remaining genes are generated, while the selected genes are \rv{set} to the specific selected values. When by \rv{setting the values of} the smallest \rv{possible} number of nodes this operation moves all possible initial conditions toward the basin of attraction of a specific attractor,  Kim {\it et al.} define \rv{this} smallest subset of nodes as the {\it control kernel} of the attractor state the system is converging to.

For consistency with our less constrained definition of what a phenotype is (subsection 2.2), we need to redefine the notion of control kernel accordingly. The generalized definition of control kernel we adopt here does not assume the forced terminal state to be a specific attractor, just a phenotype (i.e. any possible attractor compatible with the definition of that phenotype). As a result, our control kernels are significantly smaller than those found in \cite{Kim2013}. To show this explicitly, we have considered again our sample network from subsection 2.3, and the effect of \rv{setting the value of just one among genes 4-8} at a time. 

The result is shown in table 1. Forcing the expression of \rv{gene} 7 (referred to as ``7 is ON'' in the table) is enough to enlarge the basin of attraction of cell type 2 from less than 20\% to exactly 100\%. Likewise, forcing the expression of \rv{gene} 8 makes the network shift toward a phenotype characterized by the expression pattern 111, that was not even encoded in the network. Eventually, we might assume 80\% (or some other high percentage value) to be enough for the network to be considered locked in a certain cell phenotype. In this case \rv{gene} 4 would be an alternative control kernel for the cell type defined by the 111 pattern. 
This new definition of control kernel, focused on forcing the GRN into a less restrictive activation pattern than a single, specific attractor, seems to be much easier to achieve experimentally, better suited for large/realistic networks, and therefore relevant to drug treatments.

Cell type differentiation in response to extracellular signaling, is still a key aspect of our generalized approach to dynamical GRN models. Even if mainly aimed at incorporating evolutionary features (section 3), our framework entirely preserves (and sometimes facilitates) the applicability of network control theory to GRN models in explaining both developmental or drug-induced change.

\section{Conclusions}

This manuscript represents a synthesis of mathematical models of gene regulatory networks within theoretical evolutionary biology, which also accounts for development. 

Despite of their predictive power, and their possibility to mathematically reinterpret cell fates as the steady states of the abstract, non-linear dynamics of the regulatory processes they describe, dynamical models (\rv{\rv{Boolean}} models in this manuscript) of GRNs have been limited by this identification between mathematical attractors and biological phenotypes.
Evolutionary processes, like single gene and whole genome duplication, both chromosome gain and loss, horizontal gene transfer, etc., alter the size of the genome, and deprive the exact attractor-phenotype identification as a one-to-one map of its original meaning. The leading role played in evolution and speciation by the interplay between these processes and natural selection, makes current dynamical models of GRNs unfit to describe genetic network evolution.

Motivated by this limitation, we have explored the consequences of relaxing this one-to-one identification between cell types and attractors. We have shown with a simple numerical model that, redefining a cell type as a {\it collection of attractors} of a GRN, all sharing a common, characterizing property, enables retaining the main features of Kauffman's original theory, and permitting easy generalization using network control theory. It is now no longer needed for the genome to retain a fixed size for the cell type type definition to preserve its meaning.

This opens the possibility to quantitative studies of the effect that gene duplication, mutation, and natural selection have on cell types, by virtue of allowing the study of evolutionary processes within the well-established framework of dynamical systems theory.

\rv{Lastly, by showing that our generalized framework does not inhibit traditional} network controllability, \rv{we have preserved} its biological interpretation as a viable mathematical description of extracellular/epigenetic control of gene expression in an evolutionary model.

\end{multicols}



\begin{thebibliography}{99}

\bibitem{Slack2013}
J.M.W. Slack, {\it Essential Developmental Biology}, Wiley-Blackwell, Oxford 2013

\bibitem{Lynch2007}
M. Lynch, 
Nature Reviews Genetics (2007) {\bf 8} 803. 

\bibitem{Iglesias2010}
P.A. Iglesias, B.P. Ingalls, {\it Control Theory and Systems Biology}, MIT Press, Cambridge 2010 MA).

\bibitem{Stegle2015}
O. Stegle, S.A. Teichmann, J.C. Marioni, 
Nat. Rev. Genet. {\bf 16} (2015) 133.

\bibitem{Shapiro2013}
E. Shapiro, T. Biezuner, S. Linnarsson, 
Nat. Rev. Genet. {\bf 14} (2013) 618.

\bibitem{Schwartzman2015}
O. Schwartzman, A. Tanay,
Nat. Rev. Genet. {\bf 16} (2015) 716.

\bibitem{Kauffman1969}
S.A. Kauffman, 
J. Theoret. Biol. {\bf 22} (1969) 437. 

\bibitem{Delbruck1949}
M. Delbr{\"u}ck, 
Discussion. In Unit{\'e}s biologiques dou{\'e}es de continuit{\'e} g{\'e}n{\'e}tique. Paris: Editions du Centre National de la Recherche Scientifique (1949).

\bibitem{Jacob1961}
F. Jacob , J. Monod
J Mol Biol {\bf 3} (1961) 318.





\bibitem{Laubichler2017}
M.D. Laubichler, S.J. Prohaska, and  P.F. Stadler
J. Exp. Zool. (Mol Dev Evol). {\bf 330} (2018) 5.

\bibitem{Kondrashov2002}
F. A. Kondrashov, I. B. Rogozin, Y. I. Wolf, E. V. Koonin,
Genome Biol. {\bf 3} (2002) research0008

\bibitem{Conant2008}
G. C. Conant  \& K. H. Wolfe,  
Nature Rev. Genet. {\bf 9} (2008) 938.

\bibitem{Brenner1995}
S.E. Brenner, T. Hubbard, A. Murzin \& C. Chothia, 
Nature (1995) {\bf 378} 140.

\bibitem{Teichmann1998}
S.A. Teichmann, J. Park \&  C. Chothia,  
Proc. Natl. Acad. Sci. USA {\bf 95} (1998) 14658.

\bibitem{Gough2001}
J. Gough, K. Karplus, R. Hughey \& C. Chothia 
J. Mol. Biol. {\bf 313} (2001) 903.

\bibitem{Wagner1994}
A. Wagner,
{Proc. Natl. Acad. Sci.} {\bf 91} (1994) 4387.

\bibitem{Aldana2007}
M.~Aldana, E.~Balleza, S.~Kauffman, O.~Resendiz,
{\it J. Theor. Biol.} {\bf 245} (2007) 433.

\bibitem{Crombach2008}
A. Crombach, P. Hogeweg, PLoS Comp Biol {\bf 4} (2008) 7:e1000112.



\bibitem{Bonner1988}
J.T Bonner, {\it The evolution of complexity}. Princeton University Press, Princeton, New Jersey 1988.

\bibitem{Wagner1996}
G.P Wagner,
Amer. Zool., {\bf 36} (1996) 36.

\bibitem{Alberts1998}
B. Alberts, 
Cell {\bf 92} (1998) 291.

\bibitem{Hartwell1999}
L.H. Hartwell, J.J. Hopfield, S. Leibler, A.W. Murray,
Nature {\bf 402} (1999) Suppl C47.

\bibitem{Pereira-Leal2006}
J.B. Pereira-Leal, E.D. Levy, S.A. Teichmann, 
Phil. Trans. R. Soc. B {\bf 361} (2006) 507.

\bibitem{Pereira-Leal2007}
J.B. Pereira-Leal, E.D. Levy, C. Kamp, S.A. Teichmann,  
Genome Biol. {\bf 8} (2007) R51.

\bibitem{Achim2014}
K. Achim, D. Arendt,  
Curr. Opin. Genet. Dev. {\bf 27} (2014) 102.

\bibitem{Khalil2010}
A.S. Khalil, J.J. Collins, 
Nature Reviews Genetics {\bf 11} (2010) 367.

\bibitem{Hsu1993}
H. Hsu, {\it et al.},
Biophys. J. {\bf 65} (1993) 1196.

\bibitem{Stock1996}
J.B. Stock, M.G. Surette, 
in Escherichia coli and Salmonella: Cellular and Molecular Biology (1996) 1103.

\bibitem{Herskowitz1995}
I. Herskowitz, Cell {\bf 80} (1995) 187.

\bibitem{Posas1998}
F. Posas, M. Takekawa, H. Saito, 
Curr. Opin. Microbiol. {\bf 1} (1998) 175.

\bibitem{Kim2013}
J.~Kim, S.-M.~Park, K.-H.~Cho,
{\it Scientific Reports} {\bf 3} (2013) 2223.

\bibitem{Vladimir2005}
F. Vladimir,  {\it Handbook of computational molecular
biology}, University of California, Davis 2005.

\bibitem{Faure2006}
A. Faure, A. Naldi, C. Chaouiya, and D. Thieffry, 
Bioinformatics (2006) {\bf 22}
e124.

\bibitem{Garg2008}
A. Garg, A. Di Cara, I. Xenarios, {\it et al.},
Bioinformatics {\bf 24} (2008) 1917.

\bibitem{Henry2013}
A. Henry, F. Mon{\'e}ger, A. Samal, O.C. Martin,
Mol. Biosyst., {\bf 9} (2013) 1726.

\bibitem{Zhou2016}
J.X. Zhou, A. Samal, A.F d'H\'erou{\:e}l, N. D. Price, S. Huang,
Biosystems
{\bf 142} (2016) 15.

\bibitem{Ciliberti2007a}
S. Ciliberti , O.C. Martin , A. Wagner, PLoS Comp Biol {\bf 3} (2007) 2:e15.

\bibitem{Ciliberti2007b}
S. Ciliberti , O.C. Martin , A. Wagner, PNAS {\bf 104} (2007) 13591.

\bibitem{Boveri1906}
T. Boveri, 
{\it Die Organismen als Historische Wesen}.
Kgl. Universit{\"a}tsdruckerei von H. St{\"u}rtz,
1906.

\bibitem{Fumia}
H.F.~Fumi\~a, M.L.~Matins,
PLoS One {\bf 8} (2013) e69008.

\bibitem{Arendt2008}
D. Arendt, 
Nat. Rev. Genet. {\bf 9} (2008) 868.

\bibitem{Villani2011}
M. Villani, A. Barbieri, R. Serra,  
PLoS One {\bf 6} (2011) e17703.


\bibitem{Huang2012}
S. Huang,
Progress in Biophysics and Molecular Biology {\bf 110} (2012) 69. 

\bibitem{Pisco2013}
A.O. Pisco, A. Brock, J. Zhou, A. Moor, M. Mojtahedi, D. Jackson, and S. Huang, 
Nat Commun {\bf 4} (2013) 2467.

\bibitem{Attolini2009}
C.S. Attolini, F. Michor, 
Ann. N. Y. Acad. Sci. {\bf 1168} (2009) 23.

\bibitem{Diaz2012}
L.A. Diaz {\it et al.}, 
Nature {\bf 486} (2012) 537.

\end{thebibliography}
\end{document}